\documentclass[journal]{IEEEtran}
\usepackage{acronym}
\usepackage{amsthm}
\usepackage{amsmath}
\usepackage{amsmath,amsfonts}
\usepackage{amssymb}
\usepackage{algorithmic}
\usepackage[algoruled, linesnumbered]{algorithm2e}
\usepackage[english]{babel}
\usepackage{booktabs}
\usepackage{array}
\usepackage{caption}
\usepackage{cases}
\usepackage{cite}
\usepackage{cleveref}
\usepackage{colortab}
\usepackage{empheq}
\usepackage{enumitem}
\usepackage{epigraph}
\usepackage{etoolbox}
\usepackage{fixltx2e}
\usepackage[T1]{fontenc}
\usepackage{framed}
\usepackage{graphicx}
\usepackage{indentfirst}
\usepackage{latexsym}
\usepackage{lipsum}
\usepackage{makecell}
\usepackage{makeidx}
\usepackage{mathrsfs}
\usepackage{multirow}
\usepackage[cmintegrals]{newtxmath}
\usepackage{optidef}
\usepackage{pstricks}
\usepackage{ragged2e}
\usepackage[figuresright]{rotating}
\usepackage{soul}
\usepackage{stfloats}
\usepackage{subfigure}
\usepackage{tikz}
\usepackage{times}
\usepackage{textcomp}
\usepackage[normalem]{ulem}
\usepackage{url}
\usepackage{verbatim}
\usepackage{wrapfig}
\usepackage{xcolor}
\usepackage{epstopdf}
\makeindex

\SetAlCapNameFnt{\small}
\SetAlCapFnt{\small}

\DeclareGraphicsExtensions{.eps,.ps,.eps.gz,.ps.gz,.eps.Z,.png,.jpg,.bmp,.svg}
\ifCLASSINFOpdf
 
\fi
\begin{document}
%
\title{ABACUS: An Impairment Aware Joint Optimal Dynamic RMLSA in Elastic Optical Networks}
%
%
%

\author{M Jyothi Kiran, Venkatesh Chebolu, Goutam Das,~\IEEEmembership{Member,~IEEE} and Raja~Datta,~\IEEEmembership{Senior Member,~IEEE}  
\thanks{M Jyothi Kiran, R. Datta and G. Das are with Indian Institute of Technology, Kharagpur, India. V. Chebolu is with University of Cyprus, Cyprus (email:
mjyothikiran13@kgpian.iitkgp.ac.in, rajadatta@ece.iitkgp.ac.in,  gdas@gssst.iitkgp.ac.in, chebolu.venkatesh@ucy.ac.cy)}}

\maketitle

\begin{abstract}
The optimal Routing and Spectrum Assignment (RSA) presents a significant challenge in the Elastic Optical Networks (EONs). Integrating adaptive modulation formats into the RSA problem, i.e. Routing, Modulation Level and Spectrum Assignment (RMLSA),  increases allocation options and heightening complexity. The conventional RSA approach involves pre-determining fixed paths and then allocate spectrum within them separately. However, expansion of path set for optimality may not be advisable due to the substantial increase in paths with network size expansion. This paper explores a novel RMLSA, proposing a comprehensive solution addressing route determination and spectrum assignment concurrently. An objective function has been chosen and designated as ABACUS, Adaptive Balance of Average Clustering and Utilization of Spectrum. This nomenclature highlights the objective function's capability to adjust and assign significance to "average clustering" (lower fragmentation) and "spectrum utilization". Our approach involves formulating an Integer Linear Programming (ILP) model with a simple relationship between path and spectrum constraints. The model also integrates Physical Layer Impairments (PLIs) to guarantee end-to-end Quality of Transmission (QoT) for requested connections while upholding existing ones. We demonstrate that ILP can provide optimal solution for a dynamic traffic scenario within a reasonable time complexity. Towards this goal, we adopted a structured formulation approach where essential information is determined beforehand, minimizing the need for online computations. 
\end{abstract}


\IEEEpeerreviewmaketitle

\section{Introduction}

\IEEEPARstart{T}{he} traffic in the core communication network is continuously witnessing a substantial increase due to the rise in bandwidth-intensive applications such as HDTVs and Virtual or Augmented Realities. However, the available spectrum for optical communication is limited, spanning from 1530 nm to 1565 nm, amounting to approximately 4 THz. In order to meet the growing demands, it becomes crucial to efficiently utilize the available spectrum resources to accommodate the rising number of requests. Elastic Optical Networks (EONs) emerge as a potential solution to utilize spectrum flexibility beyond the capabilities of conventional optical networks operating on a fixed grid of 50 GHz [1]. By adopting EONs, the network can dynamically allocate and adjust the spectrum to optimize bandwidth utilization and meet the diverse requirements of bandwidth-intensive applications. With the transition of conventional networks towards EONs, a new challenge known as Routing and Spectrum Assignment (RSA) emerged, akin to Routing and Wavelength Assignment (RWA) in Wavelength Division Multiplexing (WDM) networks. RSA in EONs involves finding the optimal routing path and spectrum allocation for each connection [2]. RSA problem with the adaptive modulation formats, signifies a Routing, Modulation Level and Spectrum
Assignment (RMLSA) problem.
\vspace{-0.1cm}

\subsection{Static RMLSA}
 
The authors of [3] and [4] developed an ILP model for solving the RSA problem with the primary aim of minimizing total network frequency slots using a single modulation format for a connection. The formulated approach meticulously takes into account critical constraints, including non-overlapping frequencies, continuity, and contiguity. Notably, similar objective was pursued in [5] with different modulation formats for a connection. ILPs are formulated in [1], [6]-[10], each addressing distinct objective functions related to RMLSA, designed to tackle the RMLSA problem that emphasizes the inclusion of factors like nonlinear impairments, considerations pertaining to network survivability, and sophisticated strategies for bit loading. Furthermore, in [11], a node-arc ILP approach was developed specifically for static RSA that takes into consideration both routing and spectrum assignment simultaneously. Despite these advancements, the model does not provide a guarantee for Quality of Transmission (QoT).

\vspace{-0.1cm}
\subsection{Dynamic RMLSA}

In [12] and [13], the authors have addressed the RSA  challenge for dynamic traffic, incorporating considerations of distances. while notably excluding different modulations from their analyses. Further, the authors of [14] and [15] took a different approach by employing multiple modulation formats in a distance-adaptive model to solve RMLSA problem for dynamic traffic. However, these works did not account for Physical Layer Impairments  (PLIs) in their proposed solutions. The reduction of spectrum utilization was achieved through the implementation of the minimum hops concept. On the other hand, authors in [16] tackled the RSA problem with a distance-adaptive strategy using a single modulation format, concentrating on survivability. [12]-[16] solves for dynamic traffic with hueristics and in [17]-[19] the  ILP formulations for dynamic traffic are explored. In [17], the authors propose an ILP approach to address the RSA  problem for dynamic traffic by minimizing the number of hops. Notably, this solution does not take into consideration the use of multiple modulation formats or the issue of fragmentation.  In [18] and [19], the focus is exclusively on hop minimization, with no consideration given to modulation, fragmentation or PLIs.

In the aforementioned papers [1], [3]-[10], [17]-[19], a predetermined set of paths was calculated in advance for spectrum assignment. This holds true whether solving static or dynamic RSA, with or without PLIs using ILP. However, this approach incurs a drawback in that the solution of RSA is confined in identifying an optimal solution solely within the restricted set of paths which makes the solution sub-optimal. The sub-optimal solution further deviates from optimal solution with the increased node degree or in the context of larger networks as the number of possible paths between a source destination pair increases. In case of dynamic connections, the solution may deviate even more as some of the network's Frequency Slot Units (FSUs) are already in use, leaving them unavailable for upcoming connections.

In handling practical large networks, especially those with dynamic connections, ILP proves to be a workable solution. The effectiveness of this strategy lies in systematically seeking solutions explicitly designed for a specific connection request. However, as mentioned earlier, searching for the solution in a given set of predefined paths makes it sub-optimal. The process of identifying the best solution involves scrutinizing all feasible paths. This introduces an elevated level of complexity to systematically identify and precisely store all potential paths. However, it leads to the increased memory utilization and has therefore motivated us to develop an ILP where paths are not predefined. Instead, we develop a simple relation between path and spectrum assignment constraints that makes path formation and spectrum assignment mutually interdependent and provides the optimal solution for specified objective function.

This paper also takes into account PLIs when determining routes and allocating spectrum. These considerations play a pivotal role in ensuring a connection that guarantees end-to-end QoT for a Bit Error Rate (BER). However, a significant challenge arises during the computation of interference and noise power, primarily due to the unknown nature of the path through which the connection is to be established. As we are seeking for dynamic solutions and as the complete knowledge of interference and noise due to the previously established connections are known beforehand, we propose to exploit the knowledge towards the aforementioned complexity challenge to reduce the time required to achieve optimal solutions. In the process, we ensure to safeguard the end-to-end QoT for established connections all the while accommodating the allocation of newly requested connection with required QoT guarantee. Furthermore, from the literature, it is evident that most of the prior work concentrated on an objective function that either minimizes fragmentation or enhances spectrum utilization. However, it is obvious that a delicate balance is required between these two counter productive objectives. To achieve this, we have introduced a novel objective function, ABACUS, that dynamically achieves a balance between spectrum utilization and fragmentation without requiring a scaling parameter. Hence, our proposed approach adapts to the diverse requirements of the network. 

In summary, in this paper, we propose a novel RMLSA by considering the PLIs with an objective function that considers both spectrum utilization and average fragmentation without a scaling parameter and without specified predefined paths in dynamic traffic scenario. To the best of our knowledge, this comprehensive RMLSA design for dynamic traffic scenario has not been considered in the domain of EONs.

\subsection{Contributions of the paper:}
The major contributions of the paper are listed below:
\begin{itemize}
  \item We propose an ILP based 
  online solution for dynamic RMLSA which considers joint optimal Route and Spectrum Assignment and guarantees end-to-end QoT.
  \item  The proposed objective function achieves a dynamic balance between spectrum utilization and average network fragmentation.
  \item  The formulation approach is structured to determine essential information before the request is initiated, thus reducing online computations. 
  \item We have conducted extensive simulation to demonstrate that our proposal achieves a saving of 5 to 7 percent in network resources and $18 \% $ in average network fragmentation. We have also observed that consideration of PLIs ensure 100$\%$ QoT guaranteed connections while non consideration of PLIs results in 25$\%$ QoT failed connections.    
\end{itemize}

\section{Problem formulation}
In this section, we introduce a formulation of ILP for addressing the aforementioned RMLSA problem.

\subsection{Network model}
Network is modelled as $G(V, E, FSU)$ where $v_i \in V$ is a vertex (node) of the graph and $e_{i,j} \in E$ is the edge (link) between the nodes $(i,j)$ in the network having a weight equivalent to the distance between the nodes $(i,j)$ defined as $d_{i,j}$ in the network. $fsu_{i,j,k}$ corresponds to the $k^{th}$ frequency slot index in the link $(i,j)$ and is either $0$ or $1$ corresponding to whether it is unoccupied or occupied respectively. The total number of FSUs in a link is $N$. $M$ is the set of modulation formats allowed for the allocation of the spectrum where $m_i \in M$ corresponds to each modulation format. The minimum Signal to Interference Noise Ratio (SINR) required for a particular modulation format in order to maintain a desired QoT is defined and denoted as a set $SINR_{TH}=\left \{SINR_{m_1},SINR_{m_2},...,SINR_{m_M} \right \}$, where the elements $SINR_{m_1},SINR_{m_2},...,SINR_{m_M}$ corresponds to the modulation formats $m_1,m_2,...,m_M$ respectively.

\subsection{Inputs}
The inputs provided are source, $s \in V$, and destination, $d \in V$, between which a connection is being requested with a datarate, $\rho$ Gbps. This data-rate is being converted to a number of FSUs to be provided based on different modulation formats where the minimum slots required to satisfy the requested demand is given by $\rho_{m_{1}}= \left \lceil \frac{\rho}{1 \times 30}  \right \rceil, \rho_{m_{2}}=\left \lceil \frac{\rho}{2 \times 30}  \right \rceil,...,\rho_{m_{M}}=\left \lceil \frac{\rho}{M \times 30}  \right \rceil$ corresponds to the modulation formats $m_1,m_2,...,m_M$ respectively. In addition, in the path through which the preceding connection traverse, a set of FSUs ($K^{(s,d)}$) are employed using a specific modulation format $m^{(s,d)}$. Further, the accumulated SINR for a designated slot in $K^{(s,d)}$ are also included in the inputs. The notations and their definitions are listed as follows:

\begin{equation*}
    \delta_{i,j}^{s,d} \rightarrow \left\{\begin{matrix}
\text{1 if link $(i,j)$ is used in the path for allocating} \\ \text{connection between $(s,d)$ pair.} \\  \text{0 otherwise.}

\end{matrix}\right.
\end{equation*}

\begin{equation*}
    fsu_{i,j,k,m}^{s,d} \rightarrow \left\{\begin{matrix}
\text{1 if in link $(i,j)$, slot $k$ is used for } \\ \text{connection allocation between $(s,d)$ pair} \\ \text{with modulation format $m^{(s,d)}$} \\  \text{0 otherwise.}
\end{matrix}\right.
\end{equation*}

\begin{equation*}
\begin{split}
    SINR_{k}^{s,d} \rightarrow &~~\text{Signal to Interference plus Noise Ratio in the  } \\ &~~ \text{ $k^{th}$ index FSU slot for $(s,d)$ pair.}
\end{split}
\end{equation*}

\subsection{ILP Formulation}

The objective function of the proposed ILP is as follows:

minimize:

\begin{equation}
      \sum_{\forall\left ( i,j \right )\in E} \sum_{k=1}^{N}\sum_{m=1}^{M}(\log_{N} k +1)*X_{i,j,k,m}
      \label{Objective_Function_Formula}
\end{equation}

which is equivalent to 

\begin{equation*}
      \sum_{\forall\left ( i,j \right )\in E} \sum_{k=1}^{N}\sum_{m=1}^{M}(\log_{N} k )*X_{i,j,k,m} + \sum_{\forall\left ( i,j \right )\in E} \sum_{k=1}^{N}\sum_{m=1}^{M} X_{i,j,k,m}
      \end{equation*}

Here $X_{i,j,k,m}$ is a binary variable that represents the utilization of the $k^{th}$ indexed slot in link $(i,j)$ using the modulation format $m$. The objective function is designed to dynamically achieve a balance between spectrum utilization and fragmentation, adapting to the varying requirements of the network. The logarithmic parameter's significance emerges when the latter segment of the objective function attains equality across various routes. In cases where an equivalent number of FSUs is essential for connection allocation, the logarithmic function governs the decision-making process. This achievement is facilitated with careful consideration given to the selection of an appropriate base. The introduction of the logarithmic function, with a judiciously chosen base, ensures that the resultant logarithmic values remain confined within the range of $[0,1]$.  Through this process, the objective function strategically clusters sets of FSUs into lower-indexed slots, aiming to minimize the fragmentation and optimize spectrum utilization simultaneously. Consequently, we have designated this process as "ABACUS," representing Adaptive Balance of Average Clustering and Utilization of Spectrum for a requested connection. The adoption of the name "ABACUS" underscores a conscious effort to highlight the inherent adaptability, much like the dynamic movement observed in traditional Abacus beads, whether shifted back or forth based on the input for subtraction or for addition. With reference to [8],  we define the external fragmentation of a link as follows:

\begin{equation}
    1-\left ( \frac{\text{largest continuous free slots block}}{\text{total free slots}} \right )
\end{equation}

This equation holds true given the assumption that a free slot is consistently available within a link. Consequently, when no free slots are available, the fragmentation is considered as zero. The average fragmentation is then computed by considering all links throughout the network. In the next subsection, we illustrate how our proposed objective function provides a balance between fragmentation and spectrum utilization.

\subsection{Illustartion of the objective function}
\begin{figure}[!h]
    \centering
    \includegraphics[width=0.8\columnwidth, keepaspectratio]{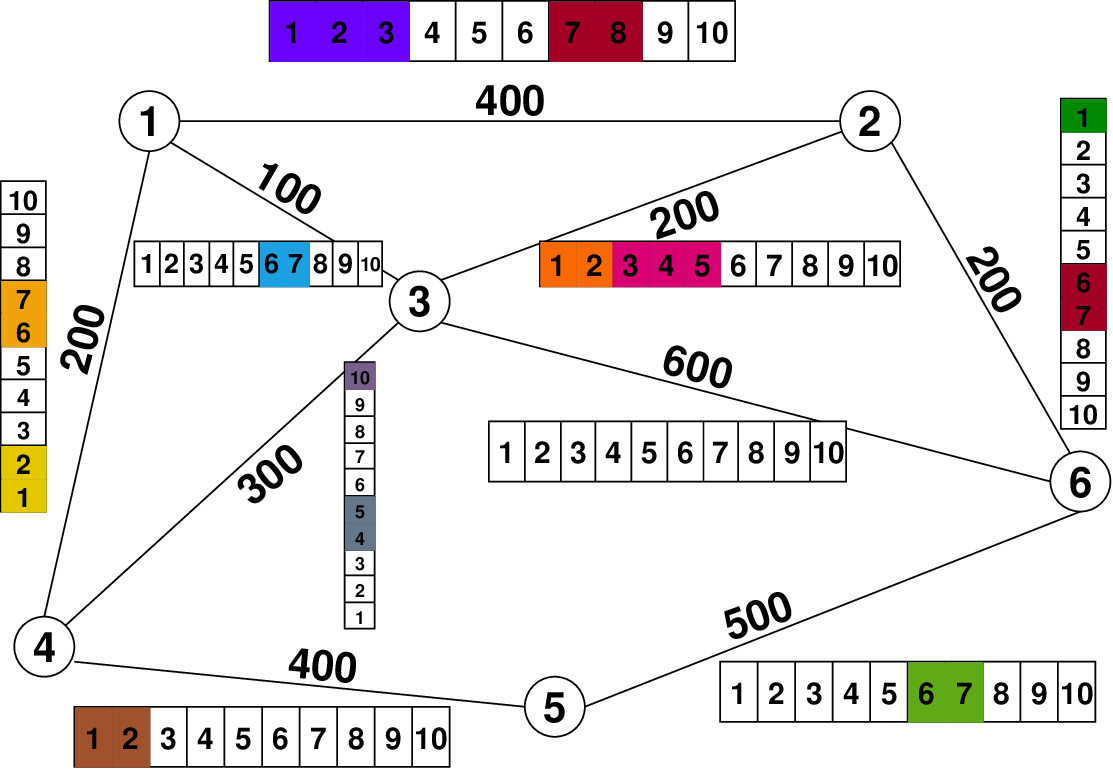}
    \caption{Six Node Network}
    \label{Six_Node_Network_2}
\end{figure}

\subsubsection{Scenario-1 : Same hops different modulation}
Let's consider a scenario wherein a connection requires allocation between node 1 and node 6, where 2 slots are required for modulation A and 4 slots for modulation B. One possible solution involves the path 1-2-6 utilizing slots 4 and 5 with modulation A, while an alternative solution is the path 1-3-6 requiring 4 slots with modulation B (the available slots here are 1, 2, 3, and 4). In this context, according to the objective function outlined in [8] and [9], the second possibility is favored due to it's lower linear sum compared to the first. However, our proposed objective function employs a logarithmic sum of FSU indices and monitors the total sum of FSUs used, playing a significant role in spectrum utilization efficiency.

\subsubsection{Scenario-2 : Same hops same modulation}
Let's consider another scenario in which a connection needs allocation between node 5 and node 3, requiring the assignment of two slots for successful establishment. Here, two potential paths, 5-6-3 and 5-4-3, both utilizing the same assumed modulation, are under consideration. Utilizing the adaptive capabilities of the defined objective function, and given the parity in the number of required slots, the logarithmic function is employed. This function facilitates allocation to the lowest indexed slots, specifically slots 1 and 2, with the objective of maintaining a consistent average fragmentation across the network—an aspect notably absent in [18], [20].

\subsubsection{Scenario - 3 : Restriction on fixed number of predefined paths}
 A third scenario may be considered in which a connection needs allocation between node 4 and node 6. In general, a fixed number of pre-determined paths, often 2 or 3, is typically considered based on distance. In that case, the path set would be 4-1-2-6, 4-3-1-2-6 and 4-3-2-6. However, sticking to such a practice may result in overlooking two potentially advantageous paths, namely 4-5-6 and 4-3-6, both characterized by fewer hops compared to the initial three paths. This oversight can directly influence spectrum utilization.

Henceforth, we shall refer to the objective function presented in [8] as RMLSA-KSP(2) and RMLSA-KSP(3) for k = 2 and 3 paths respectively. The objective function in [8] with the ILP formulation in our paper to be referred as JO-RMLSA. Additionally, the objective function with ILP formulation proposed in our paper  will be referred to as ABACUS-RMLSA.

\subsection{Input Variables}

The parameters essential for the evaluation of the objective function are specified as follows:

\subsubsection{Binary variables}

\begin{equation*}
    X_{i,j} \rightarrow \left\{\begin{matrix}
\text{1 if link $(i,j)$ is used in the path.} \\  \text{0 otherwise.}

\end{matrix}\right.
\end{equation*}

\begin{equation*}
    X_{i,j,k,m} \rightarrow \left\{\begin{matrix}
\text{1 if in link $(i,j)$, modulation format} \\ \text{m is used in slot k.} \\ \text{0 otherwise.}
\end{matrix}\right.
\end{equation*}

\begin{equation*}
    Z_{i,j,k,m} \rightarrow \left\{\begin{matrix}
\text{1 if slot k of link $(i,j)$ is the first slot} \\ \text {for modulation format m. } \\  \text{0 otherwise}

\end{matrix}\right.
\end{equation*}

\subsection{Constraints for Routing and Spectrum Assignment}
The constraints necessary for joint path and spectrum allocation are listed below:

\subsubsection{Relation between variables}

\begin{equation*}
    Z_{i,j,k,m} \leq X_{i,j,k,m} \leq X_{i,j};    
\end{equation*}

\begin{equation}
   \hspace{2cm} \forall (i,j) \in E ,\forall k \in N,\forall m \in M
   \label{Relation b_w variables}
\end{equation}

Equation (\ref{Relation b_w variables}) establishes a relationship between path variables and spectrum assignment variables to ensure joint optimization.

\subsubsection{Path formation constraints}

\begin{equation}
    \sum_{j\in V-\left \{ s \right \}} X_{s,j} = 1;
    \label{source to first node}
\end{equation}

\begin{equation}
    \sum_{i\in V-\left \{ d \right \}} X_{i,d} = 1;
    \label{last but one to destination node}
\end{equation}

\begin{equation}
    \sum_{i\in V-\left \{ d \right \}}^{} X_{i,j} = \sum_{k\in V-\left \{ s \right \}}^{} X_{j,k}; ~~~~  \forall j \in V-\left \{ s,d \right \}
    \label{path continuity}
\end{equation}

\begin{equation}
    \sum_{j\in V-\left \{ i \right \}}^{} X_{i,j} \leq 1; ~~~~ \forall i \in V-\left \{ d \right \}
    \label{one outgoing path}
\end{equation}

\begin{equation}
    \sum_{i\in V-\left \{ j \right \}}^{} X_{i,j} \leq 1;   ~~~~~ \forall j \in V-\left \{ s \right \}
    \label{one incoming path}
\end{equation}

\begin{equation}
    X_{i,j} + X _{j,i} \leq 1; ~~~~~ \forall i \in V, \forall j \in V, i \neq j
    \label{one path}
\end{equation}

Equations (\ref{source to first node})-(\ref{path continuity}) establish a continuous path between a given source and destination, while equations (\ref{one outgoing path})-(\ref{one path}) guarantee the existence of a singular path.

\subsubsection{Spectrum assignment initialization}

\begin{equation}
    \sum_{j = 1}^{V} \sum_{k = 1}^{N} \sum_{m = 1}^{M} Z_{s,j,k,m} = 1
    \label{spectrum assignment initialization}
\end{equation}

Owing to the interdependency established in equation  (\ref{Relation b_w variables}) and the path formation determined by equations (\ref{source to first node})-(\ref{path continuity}), equation (\ref{spectrum assignment initialization}) ensures that only a single starting slot can be populated within the selected path. 

\subsubsection{Spectrum contiguity and continuity constraints}

\begin{equation}
        \sum_{t=k}^{k+\rho_{m} -1} X_{i,j,t,m} \geq \rho_{m} \ast Z_{i,j,k,m};   
    \label{Contiguity}
\end{equation}

\vspace{-0.5cm}
\begin{equation*}
    \hspace{2.5cm} \forall \left ( i,j \right ) \in E,  \forall k \in N,~  \forall m \in M 
\end{equation*}

\begin{equation}
        \sum_{k=1}^{N} X_{i,j,k,m} \leq \rho_{m};       
    \label{Restricting_max_slots}
\end{equation}

\vspace{-0.3cm}
\begin{equation*}
   \hspace{2.5cm} \forall \left ( i,j \right )\in E, ~\forall m \in M
\end{equation*}

\begin{equation}
    \begin{split}
        \sum_{i\in V-\left \{ d \right \}}^{} Z_{i,j,k,m}  = \sum_{q\in V-\left \{ s \right \}}^{} Z_{j,q,k,m}; & ~~ \forall j \in V- \left \{ s,d \right \},\\
        & \forall k \in N, ~ \forall m \in M 
    \end{split} 
    \label{Continuity_Z}
\end{equation}

Equations (\ref{Contiguity}) and (\ref{Restricting_max_slots}) ensures the selection of the requisite number of slots in a contiguous order, while equation (\ref{Continuity_Z}) guarantees the preservation of path continuity corresponding to the chosen modulation format.

\subsubsection{Non-overlapping and link capacity constraints}

\begin{equation}
    \begin{split}
        fsu_{i,j,k} + X_{i,j,k} \leq  1; & ~~ \forall \left ( i,j \right )\in E, \\ &  ~~~\forall  k\in N 
    \end{split}
    \label{Non overlapping}
\end{equation}

where
\begin{equation*}
    X_{i,j,k} = \sum_{m=1}^{M}X_{i,j,k,m}; ~\forall (i,j) \in E,~ \forall k \in N    
\end{equation*}

Equation (\ref{Non overlapping}) guarantees that the slots occupied by preceding connections remain inaccessible to the current allocation and equation (\ref{Link_Capacity}) ensures that the link capacity constraint is not violated.

\begin{equation}
    \sum_{k=1}^{N} \sum_{m=1}^{M} X_{i,j,k,m} \leq  N;  \hspace{1cm} \forall \left ( i,j \right )\in E
    \label{Link_Capacity}
\end{equation}

\subsection{Additional constraint without PLIs}
\begin{equation}
   \sum_{\left ( i,j \right ) \in E} \sum_{k=1}^{N} Z_{i,j,k,m}\ast  d_{i,j}\leq D_{m}; ~~~ \forall m \in M 
   \label{distance_aware_constraint}
\end{equation}

Equation (\ref{distance_aware_constraint}) guarantees the maximum optical reach permitted by a specific modulation where maximum optical reach is given as $D_1,D_2,...,D_M$ corresponding to $m_1,m_2,...,m_M$.

\subsection{Additional constraints with PLIs}
In this section, we have structured our ILP formulation so that majority of the calculations can be performed offline. 

\subsubsection{In-band Crosstalk}
In-band crosstalk refers to the interference that occur when two or more optical channels or signals share identical frequency slots and the nodes. It has the potential to deteriorate the quality of transmitted signal and consequently elevate the occurrence of bit errors, thereby exerting an adverse impact on the overall network performance. The detailed description of in-band crosstalk was discussed in [21].

Therefore, it becomes imperative to account for the crosstalk power within the transmission path in which we are allocating a connection. Given that the specific path is not known apriori, it becomes necessary to consider crosstalk powers at all nodes. However, it is feasible to perform this calculation prior to commencing the ILP optimization process as all the prior connections are available at that stage. The parameters for crosstalk calculation listed below:

\begin{equation*}
\begin{split}
    Pxt_{i,j,k} \rightarrow & ~~\text{Crosstalk power at node $i$ in link $(i,j)$} \\ &\text{for the FSU `k'.}
\end{split}
\end{equation*}

\begin{equation*}
\begin{split}
    Pxt_{total,k} \rightarrow &~~\text{Total cross-talk power accumulated for} \\ &\text{ the connection in the slot `k'.}
\end{split}
\end{equation*}

 The equations (\ref{Crosstalk_node_wise}) and (\ref{cumulative_crosstalk}) provide the crosstalk power at individual nodes as well as the cumulative crosstalk power along the selected path.

 \begin{equation}
     Pxt_{i,j,k} = \sum_{j^{'}=1}^{V-\left \{ i,j \right \}} P_{x} \times fsu_{j^{'},i,k};    
     \label{Crosstalk_node_wise}
\end{equation}

\vspace{-0.6cm}
\begin{equation*}
    \hspace{4cm} \forall \left ( i,j \right ) \in E , \forall k \in N
\end{equation*}

where, $P_{x} = P_{r}\times C_{x}$ represents the power of the interfering signal at a given node. $C_{x}$ denotes the crosstalk factor, and $P_{r}$ stands for the received power.  Importantly, $P_{r}$ remains constant regardless of modulation format and slot under the assumption of same power across all slots. This assumption makes the PLI calculation independent of $m$; significantly reducing the overall complexity. As equation (\ref{Crosstalk_node_wise}) does not include any ILP variables, it can be precalculated offline.

\begin{equation}
    Pxt_{total,k} = \sum_{\left ( i,j \right ) \in E }^{} Pxt_{i,j,k} \times X_{i,j}; \hspace{0.5cm} \forall k \in N 
    \label{cumulative_crosstalk}
\end{equation}

In the analysis, we compute the cumulative crosstalk power for each slot along the selected path. Additionally, it is essential to validate the increment in crosstalk power resulting from the current connection to ensure that it remains within acceptable limits to safeguard the QoT for the existing connections. The crosstalk increment introduced to each source-destination pair is expressed in the subsequent equation (\ref{Previous_connections_Crosstalk}). Crosstalk for previous connections were calculated only for those frequency slots which were previously occupied.

\begin{equation}
    Pxt_{extra,k}^{s,d} = \sum_{\left ( i,j \right ) \in E }^{} \sum_{j^{'}=1}^{V-\left \{ i,j \right \}} P_{x} \times fsu_{i,j,k}^{s,d} \times X_{j',i,k};
    \label{Previous_connections_Crosstalk}
\end{equation}

\vspace{-0.3cm}
\begin{equation*}
    \hspace{3cm} \forall k \in K^{(s,d)}, ~~ \forall (s,d) pairs
\end{equation*}

\subsubsection{Nonlinear Interference constraints}
It is observed that Nonlinear Interference (NLI) encompass adverse effects on signal quality similar to the in-band crosstalk. The meticulous assessment of is important for evaluating QoT and upholding the dependability. Subsequent sections will delve into the consideration of self-channel interference and cross-channel interferences. We now employ the Gaussian noise model as detailed in [8] to calculate the impact of NLI. The parameters governing the NLI constraints are subsequently defined as follows:

\begin{equation*}
\begin{split}
Pnli_{total,k} \rightarrow &~~\text{Total nonlinear interference power } \\ &\text{accumulated for the connection in  slot `k'.}
\end{split}
\end{equation*}

\begin{equation*}
\begin{split}
    Pnli_{extra,k}^{s,d} \rightarrow \text{ Extra nonlinear interference added to $(s,d)$  } \\ \text{pair because of assigning the current connection.}
\end{split}
\end{equation*}

The NLI power for a slot within a given link is computed by assessing the cumulative interference contributions from Self-channel interference $(Pnli_{SCI})$, interference from other established connections $(Pnli_{XCI-oc})$ and by the other slots within the same connection $(Pnli_{XCI-pc})$.

\begin{equation}
    Pnli_{SCI} = \Omega \Delta f G^{3}\left ( f \right ) \ln \left |\frac{\pi ^{2}\beta _{2}\left ( \Delta f \right )^{2}}{\alpha }  \right | 
    \label{SCI-Equation-seperate}
\end{equation}

\begin{equation}
    Pnli_{XCI-oc_{i,j,k}} = \Omega \Delta fG\left ( f \right ) \sum_{k^{'}=1}^{N} G^{2}\left ( f^{'} \right ) \ln\left | \mu  \right | fsu_{i,j,k};
    \label{XCI-equation-seperate}
\end{equation}

\vspace{-1cm}
\begin{equation*}
    \hspace{4cm} \forall(i,j) \in E, \forall k \in N
\end{equation*}

\begin{equation}
    Pnli_{XCI-pc_{i,j,k}} = \Omega \Delta fG\left ( f \right ) \sum_{k^{'}=1}^{N} G^{2}\left ( f^{'} \right ) \ln\left | \mu  \right | X_{i,j,k};
\end{equation}

\vspace{-0.5cm}
\begin{equation*}
    \hspace{4cm} \forall(i,j) \in E, \forall k \in N
\end{equation*}

\begin{equation*}
\begin{split}
    Pnli_{total,k} = & \sum_{\left ( i,j \right ) \in E } \left ( Pnli_{SCI}X_{i,j} + Pnli_{XCI-oc_{i,j,k}} X_{i,j} + \right. \\   & \left.\hspace{0.8cm} Pnli_{XCI-pc_{i,j,k}} \right )  \times \left \lceil \frac{d_{ i,j }}{span Length} \right \rceil;
\end{split}
\end{equation*}

\begin{equation}
     \hspace{5cm} \forall k \in N
    \label{NLI_equation_in_link_K}
\end{equation}
where, 
\begin{equation*}
\Omega  = \frac{3\gamma ^{2}}{2\pi \alpha \left | \beta_{2} \right |} ~ \text{and}~ \mu  = \frac{\left | k - k^{'} \right |+ \frac{\Delta f}{2}}{\left | k - k^{'} \right |- \frac{\Delta f}{2}}
\end{equation*}

where $G(f)$ is the power spectral density of the signal having bandwidth as $\Delta f$ with center frequency $f$. $\beta_{2}$ is the fiber dispersion, $\gamma$ is the fiber nonlinear coefficient and $\alpha$ is the power attenuation. Further, $\Delta f^{'}$ is the bandwidth of the other signal having a center frequency $f^{'}$. As equations (\ref{SCI-Equation-seperate}) and (\ref{XCI-equation-seperate}) does not include any ILP variables, those can be precalculated offline.

Similar to the crosstalk phenomena, the nonlinear impairments of the preceding connections are also influenced by the allocation of FSUs from the connection currently being assigned and is noted in the equation (\ref{NLI_previous_connection}).

\begin{equation*}
\begin{split}
Pnli_{extra,k}^{s,d} & = \sum_{\left ( i,j \right ) \in E}^{}\Omega \Delta fG\left ( f \right ) \sum_{k^{'}}^{} G^{2}\left ( f^{'} \right ) \ln\left | \mu  \right | \\ & \times  X_{i,j,k} \times \delta_{i,j}^{s,d} \times \left \lceil \frac{d_{ i,j }}{span Length} \right \rceil;
\end{split}
\end{equation*}

\vspace{-0.4cm}
\begin{equation}
    \hspace{3.5cm} \forall k \in K^{(s,d)}, ~~ \forall (s,d) pairs
    \label{NLI_previous_connection}
\end{equation}

\begin{table}[h]
    \centering
    \begin{tabular}{|c|c|}
        \hline
        \textbf{Parameter} & \textbf{Value} \\
        \hline
        LO power $(P_{lo})$, Received power $(P_{r})$  & 0 dBm,-12 dBm \\
        \hline
        Responsivity $(R_{a})$, Operating frequency $(f_{c})$ & 0.7 A/W, 193.1 THz \\
        \hline
        Spontaneous Emission factor $(n_{sp})$  & 2 \\
        \hline
        Switch through loss (dB) assuming  & 3 $\left \lceil \log _{2} Q \right \rceil + L_{WSS}$ \\
        broadcast and select architecture & $Q: inputs/outputs$ \\
        \hline
        Fiber attenuation $(\alpha)$, WSS loss$(L_{WSS})$ & 0.2 dB/km,2 dB \\
        \hline
        Tap loss $(L_{tap})$ & 1 dB \\
        \hline
        EDFA spacing and fiber span   (L) & 80 km \\
        \hline
        Input EDFA gain  $(G_{in})$ & 18 dB \\
        \hline
        Output EDFA gain $(G_{out})$ in & 5 dB at node 7 and 10, \\
        14 Node NSFNET (fig. \ref{NSFNET_Figure}) & 8 dB elsewhere \\
        \hline
        Crosstalk factor $(C_{x})$ &  -40 dB \\
        \hline
        Nonlinear coefficient $(\gamma)$ & 1.33 $W^{-1}km^{-1}$ \\
        \hline
        Fiber dispersion $(\beta_{2})$ & -21.7 $ps^{2}/km$ \\
        \hline
        Electrical Bandwidth $(B_{e})$ & 7 GHz \\
        \hline
        Planck's Constant (h) & 6.62 $\times$ 10$^{-34}$ J.s \\
        \hline
        Slot Width $(\Delta f)$  [9] & 37.5 GHz \\
        \hline
    \end{tabular}
    \caption{Simulation Parameters}
    \label{tab:mytable}
\end{table}

\subsubsection{Local oscillator (LO)-ASE beat
noise variance}
The variance of Local Oscillator (LO) - Amplified Spontaneous Emission (ASE) beat noise is contingent upon the path's length and is given in [8] as

\begin{equation}
    \sigma ^{2}_{lo-sp} = \zeta \left ( T\left ( G_{in}-1 \right ) +\sum_{i=1}^{V} \left (X_{i} \times G_{out}\left ( i \right ) -1 \right ) \right )
\end{equation}

where $T$ is the total number of EDFAs in the path, $G_{in}$ is the input gain of EDFA and $G_{out}\left ( i \right )$ is the output EDFA gain at nodes $i = 1,2,.. H$, which are the nodes through which the connection is allocated and by [22], $G_{out}\left ( i \right ) \geq 3 \left \lceil \log _{2} Q\left ( i \right ) \right \rceil + L_{WSS} $ dB where $Q\left(i\right)$ is the fiber inputs/outputs at node $i$. 

\begin{equation*}
    T = \frac{\sum_{\left ( i,j \right )\in E}^{}X_{i,j}*d_{i,j}}{L} ,  ~~  \zeta = \frac{R_{a}^{2}}{2}P_{lo}2n_{sp}hf_{c}B_{e},   
\end{equation*}

\begin{equation*}
   X_{i} = \sum_{j=1}^{V} X_{i,j} ~~\forall i \in V
\end{equation*}

Refer Table-I for each term.
\subsubsection{Necessary condition for QoT guarantee}
To uphold the QoT, it is imperative to ensure that the SINR computed for a specific time slot exceeds the prescribed minimum threshold, as expressed in equation (\ref{SINR_threshold_equation}).
\begin{equation}
    \begin{split}
        SINR \geq SINR_{th} & ~~ \implies \frac{P_{ch}}{P_{N}+P_{I}} \geq SINR_{th}  \\&~~ \implies \frac{P_{N}+P_{I}}{P_{ch}} \leq \frac{1}{SINR_{th}}
        \label{SINR_threshold_equation}
    \end{split}
\end{equation}

where, $P_{ch}$ : Coherently received signal power given as $((R_{a}^2)/2 \times P_{lo} \times P_{r})$, $P_{I}$ : Interference power, $P_{N}$ : Noise power and $SINR_{th}$ : Threshold SINR value to maintain a particular BER.

Different modulations are associated with varied SINR thresholds. Therefore, it is essential to ensure that the each allocated slot meets the requisite SINR threshold as shown in equation (\ref{SNR_Final_equation}).


\begin{equation}
\begin{split}
    \frac{\sigma _{lo-sp}^{2} + \sigma _{lo-xt}^{2|k} + \sigma _{lo-nli}^{2|k}}{P_{ch}} + NSIS*&X_{i,j,k,m} \leq NSIS + SIS_{m}; \\ & \forall ~  i \in V, ~~\forall ~j \in V, \\ & \forall ~k \in N, \forall ~ m \in M
    \label{SNR_Final_equation}
\end{split}
\end{equation}

where ,
\begin{equation*}
    \sigma _{lo-xt}^{2|k} = (R_{a}^2)/2 \times P_{lo} \times Pxt_{total,k},
\end{equation*}
 
\begin{equation*}
    \sigma _{lo-nli}^{2|k} = (R_{a}^2)/2 \times P_{lo} \times Pnli_{total,k},
\end{equation*}

\noindent are the variances of crosstalk power and NLI power. $SIS_m$ represents the value of inverse of SINR thresholds for each modulation format $m \in M$ and NSIS is any value greater than all the $SIS_{m}$ values.

\begin{equation}
    \frac{1}{SINR_{k}^{s,d}}+\frac{\sigma _{lo-xt}^{2|\left ( s,d,k \right )}+\sigma _{lo-nli}^{2|\left ( s,d,k \right )}}{P_{ch}} \leq  SIS_{m};
    \label{SNR_Final_equation_previous_connections}
\end{equation}

\vspace{-0.4cm}
\begin{equation*}
    \hspace{2.5cm} \forall k \in K^{(s,d)}, ~~ \forall (s,d) pairs,  m = m^{(s,d)}
\end{equation*}

Equation (\ref{SNR_Final_equation_previous_connections}) ensures that after the allocation of the current connection, the SINR of all existing connections must exceed the SINR thresholds corresponding to their associated modulation formats.
 
\section{Results and Discussion}

\begin{figure}[h]
    \centering
    \includegraphics[width=0.8\columnwidth, keepaspectratio]{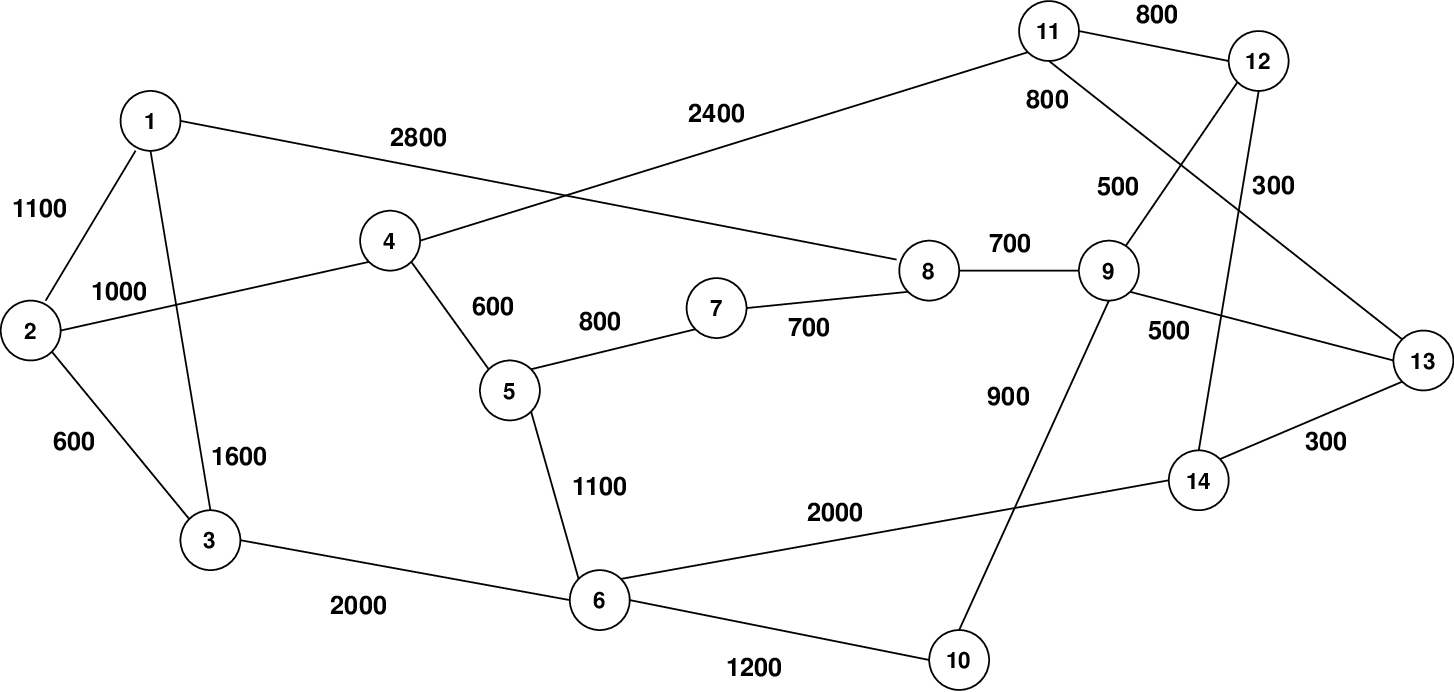}
    \caption{14 Node NSFNET}
    \label{NSFNET_Figure}
\end{figure}

\begin{figure*}[h]
    \centering
    \subfigure[]{\includegraphics[width=0.3\textwidth,keepaspectratio]{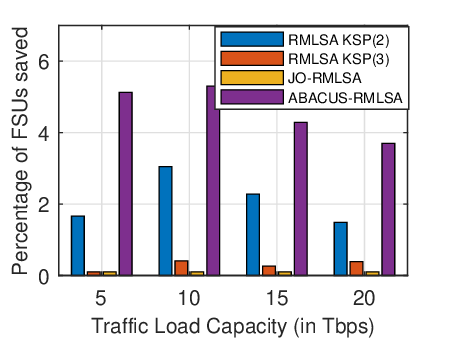}}
    \subfigure[]{\includegraphics[width=0.3\textwidth,keepaspectratio]{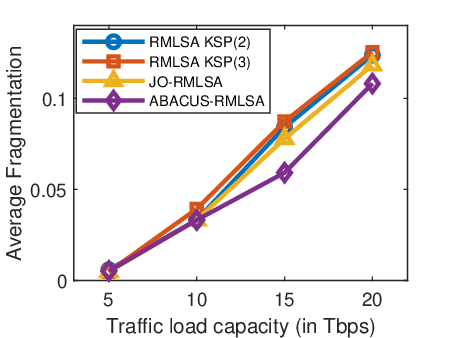}}
    \subfigure[]{\includegraphics[width=0.3\textwidth,keepaspectratio]{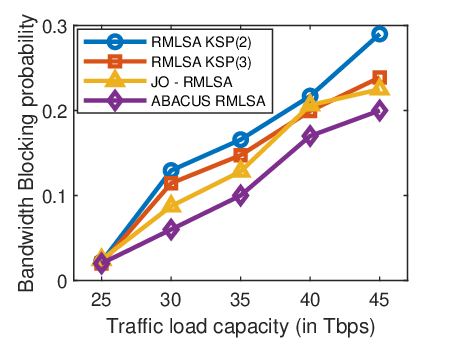}}
    \subfigure[]{\includegraphics[width=0.3\textwidth,keepaspectratio]{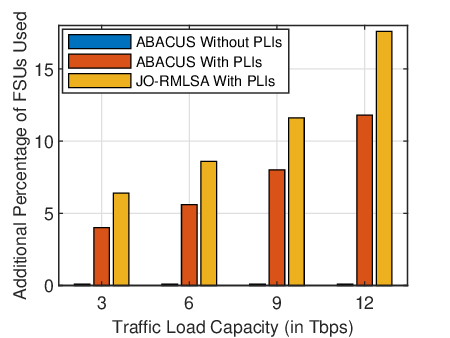}}
    \subfigure[]{\includegraphics[width=0.3\textwidth,keepaspectratio]{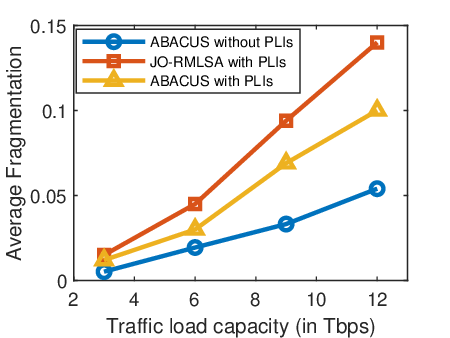}}
    \subfigure[]{\includegraphics[width=0.3\textwidth,keepaspectratio]{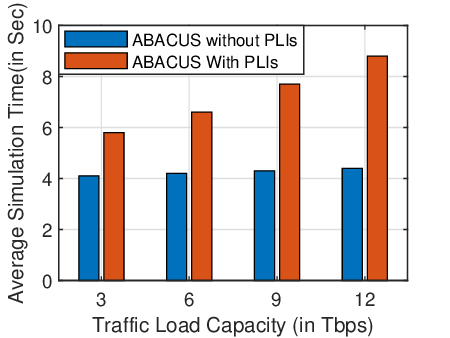}}
    
    \caption{Performance analysis in 14-node NSFNET for different Traffic Loads vs.(a) Percentage of FSUs saved, (b) Average Fragmentation, (c) Bandwidth Blocking Probability, (d) Additional Percentage of FSUs used, (e) Average Fragmentation with PLIs, (f) Average Simulation Time }
    \label{14_Node_Simulation_Results}
\end{figure*}

In this section, a comparative analysis is conducted between our proposed ILP formulation and the pre-existing ILP models. Details of the simulation parameters are available in Table I. The demand requests, originating from source-destination (s-d) pairs, are generated with a uniform distribution. Simulations have been conducted for a 14-node NSFNET illustrated in Fig. \ref{NSFNET_Figure},  employing varying data rates within the range of 70 Gbps to 700 Gbps characterized by a uniform distribution. We have considered N =110 slots in a link. In this analysis, the modulation formats considered are BPSK, 4-QAM, 8-QAM, and 16-QAM. The SINR thresholds corresponding to a BER of $10^{-9}$ are 12.6 dB, 15.6 dB, 19.2 dB, and 22.4 dB for BPSK, 4-QAM, 8-QAM, and 16-QAM, respectively [8]. We used IBM CPLEX Optimization studio for ILP and MATLAB R2023b for generating text files. The processor used for simulations is 12th Gen Intel(R) Core(TM) i5-1235U.

\subsection{Performance Analysis of a 14-Node NSFNET}
\subsubsection{Without PLIs}
For NSFNET, we generate a comprehensive set of simulation results averaging over 15 distinct statistically identical input sets. We observed that, up to an aggregate data rate of 20 Tbps load, none of the connections experienced blocking. Following this threshold, calculations were performed for the average number of FSUs saved and the average fragmentation. Subsequently, the average bandwidth blocking probability was determined up to 45 Tbps and all with no PLIs considered.

In the context of spectrum utilization, the proposed formulation consistently saves 5 to 6 percent of FSUs, along with a minimum of 18 percent reduction in the average fragmentation when compared to RMLSA-KSP(2), RMLSA-KSP(3), JO-RMLSA under similar simulation scenario presented in Fig. \ref{14_Node_Simulation_Results}.(a) and Fig. \ref{14_Node_Simulation_Results}.(b) respectively. This outcome is attributed to the efficient utilization of available spectrum within the proposed objective function, which effectively balances average network fragmentation. Fig. \ref{14_Node_Simulation_Results}.(c) presents a comparative analysis from 25 Tbps to 45 Tbps, indicating the proposed objective has the lowest bandwidth blocking probability with a minimum of 22 percent at 45 Tbps traffic load capacity due to lower FSU occupancy and reduced average fragmentation.

\subsubsection{With PLIs}
\begin{figure}
    \centering
    \includegraphics[width=0.7\columnwidth, keepaspectratio]{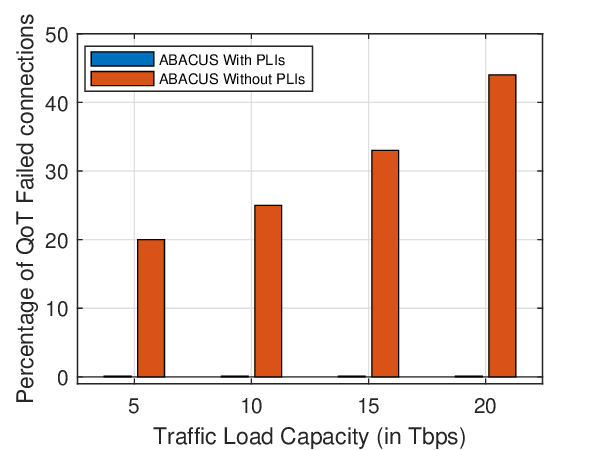}
    \caption{Traffic Load vs Percentage of QoT Failed Connections}
    \label{QoT_Result}
\end{figure}

As previously mentioned, to guarantee the QoT, consideration of PLIs plays a significant role. To validate this assertion, simulations were conducted using same traffic considered in the preceding section. Three distinct scenarios were examined: "ABACUS without PLIs," denoting the ILP with ABACUS-RMLSA without PLI constraints; "ABACUS with PLIs," representing the ILP with ABACUS-RMLSA with PLI constraints; and "JO-RMLSA with PLIs," wherein JO-RMLSA is considered with PLIs. Fig. \ref{14_Node_Simulation_Results}.(d) and Fig. \ref{14_Node_Simulation_Results}.(e) demonstrates that the inclusion of PLIs requires an additional 6 percent of FSUs and leads to a 26 percent increase in the average network fragmentation in order to accommodate the same set of requests. However, this allocation strategy ensures the QoT, as presented in Fig. \ref{QoT_Result}, which demonstrates that approximately 25 percent of requests allocated without considering PLIs fail to meet the QoT criteria at a traffic load of 6 Tbps. It is significant to note that there is still a saving of 35 percent in spectrum and a 21.5 percent in average network fragmentation with ABACUS-RMLSA when compared with the JO-RMLSA with PLIs. This is due to the adaptive characteristics of the objective function. Notably, the majority of computations are conducted offline to speed up the online solution process, resulting in an average processing time of approximately 5.8 seconds at 3 Tbps load and 8.9 seconds during simulation at 12 Tbps load for ABACUS-RMLSA with PLIs. This extended processing time is attributed to the additional calculations required to ensure the QoT for existing connections. Conversely, the ABACUS-RMLSA without PLIs maintains a consistent simulation time of approximately 4.1 seconds across all loads, as it does not incorporate impairment parameters into its calculations. The analysis corresponding to simulation timings are presented in Fig. \ref{14_Node_Simulation_Results}.(f).

\section{Conclusion}
This paper propose a novel RMLSA that considers PLIs with an objective function that balances both spectrum utilization and fragmentation without a scaling parameter. The proposed RMLSA also jointly optimizes route and spectrum selection in dynamic traffic scenario with adaptive modulation formats. Our proposal employs ILP to concurrently solve optimal routes and spectrum allocation by establishing a direct relationship between path variables and spectrum assignment variables. This formulation also ensures the end-to-end QoT for both requested and established connections, accounting for PLIs.

From a performance perspective, our proposed approach consistently achieves savings of 5 to 7 percent in network resources, accompanied by a reduction of approximately $18 \% $ in average network fragmentation. This improvement facilitates the allocation of more connections, thereby lowering the probability of bandwidth blocking by around $22 \%$. It is further observed that approximately $25\%$ of connections experience failure in ensuring the end-to-end QoT when PLIs are not taken into account during connection allocation. In terms of timing considerations, in a practical scenario involving a 14-node NSFNET with 110 FSUs, the average time required to successfully allocate a connection is around 8 seconds at 12 Tbps.

{\fontsize{8}{10}\selectfont 
\section*{References}
\begin{enumerate}
    \item K. Christodoulopoulos, I. Tomkos, and E. A. Varvarigos, “Routing and spectrum allocation in ofdm-based optical networks with elastic bandwidth allocation,” in 2010 IEEE Global Telecommunications Conference GLOBECOM 2010. IEEE, 2010, pp. 1–6.
    \item Y. Wang, X. Cao, and Y. Pan, “A study of the routing and spectrum allocation in spectrum-sliced elastic optical path networks,” in 2011 Proceedings Ieee Infocom. IEEE, 2011, pp. 1503–1511.
    \item J. H. Capucho and L. C. Resendo, “Ilp model and effective genetic algorithm for routing and spectrum allocation in elastic optical networks,” in 2013 SBMO/IEEE MTT-S International Microwave \& Optoelectronics Conference (IMOC). IEEE, 2013, pp. 1–5.
    \item M. Klinkowski and K. Walkowiak, “Routing and spectrum assignment in spectrum sliced elastic optical path network,” IEEE Communications Letters, vol. 15, no. 8, pp. 884–886, 2011.
    \item K. Christodoulopoulos, I. Tomkos, and E. A. Varvarigos, “Elastic bandwidth allocation in flexible ofdm-based optical networks,” Journal of Lightwave Technology, vol. 29, no. 9, pp. 1354–1366, 2011.
    \item Y. Miyagawa, Y. Watanabe, M. Shigeno, K. Ishii, A. Takefusa, and A. Yoshise, “Bounds for two static optimization problems on routing and spectrum allocation of anycasting,” Optical switching and networking, vol. 31, pp. 144–161, 2019.
    \item J. Wang, H. Xuan, Y. Wang, Y. Yang, and S. Liu, “Optimization model and algorithm for routing and spectrum assignment in elastic optical networks,” in 2018 14th International Conference on Computational Intelligence and Security (CIS). IEEE, 2018, pp. 306–310.
    \item S. Behera, A. Deb, G. Das, and B. Mukherjee, “Impairment aware routing, bit loading, and spectrum allocation in elastic optical networks,” Journal of Lightwave Technology, vol. 37, no. 13, pp. 3009–3020, 2019.
    \item V. Chebolu, S. Behera, and G. Das, “Robust rmsa design for shared backup path protection based eon against nonlinear impairments,” Journal of Optical Communications and Networking, vol. 14, no. 8, pp. 640–653, 2022.
    \item D. Adhikari, R. Datta, and D. Datta, “Design methodologies for survivable elastic optical networks with guardband-constrained spectral allocation,” in 2018 3rd International Conference on Microwave and Photonics (ICMAP). IEEE, 2018, pp. 1–6.
    \item A. Cai, G. Shen, L. Peng, and M. Zukerman, “Novel node-arc model and multiiteration heuristics for static routing and spectrum assignment in elastic optical networks,” Journal of Lightwave Technology, vol. 31, no. 21, pp. 3402–3413, 2013.
    \item X. Wan, L. Wang, N. Hua, H. Zhang, and X. Zheng, “Dynamic routing and spectrum assignment in flexible optical path networks,” in 2011 Optical Fiber Communication Conference and Exposition and the National Fiber Optic Engineers Conference. IEEE, 2011, pp. 1–3.
    \item T. Takagi, H. Hasegawa, K.-i. Sato, Y. Sone, B. Kozicki, A. Hirano, and M. Jinno, “Dynamic routing and frequency slot assignment for elastic optical path networks that adopt distance adaptive modulation,” in Optical Fiber Communication Conference. Optica Publishing Group, 2011, p. OTuI7.
    \item L. R. Costa and A. C. Drummond, “Dynamic multi-modulation allocation scheme for elastic optical networks,” in 2021 International Conference on COMmunication Systems \& NETworkS (COMSNETS). IEEE, 2021, pp. 596–604.
    \item A. N. Khan, “Online routing, distance-adaptive modulation, and spectrum allocation for dynamic traffic in elastic optical networks,” Optical Fiber Technology, vol. 53, p. 102026, 2019.
    \item C. Wang, G. Shen, and S. K. Bose, “Distance adaptive dynamic routing and spectrum allocation in elastic optical networks with shared backup path protection,” Journal of Lightwave Technology, vol. 33, no. 14, pp. 2955–2964, 2015.
    \item J. Zhang, P. Miao, and F. Zhang, “On optimal routing and spectrum allocation in elastic optical networks,” in 2023 2nd International Conference on Big Data, Information and Computer Network (BDICN). IEEE, 2023, pp. 284–287.
    \item X. Chen, S. Zhu, D. Chen, S. Hu, C. Li, and Z. Zhu, “On efficient protection design for dynamic multipath provisioning in elastic optical networks,” in 2015 International Conference on Optical Network Design and Modeling (ONDM). IEEE, 2015, pp. 251–256.
    \item L. Ruan and Y. Zheng, “Dynamic survivable multipath routing and spectrum allocation in ofdm-based flexible optical networks,” Journal of Optical Communications and Networking, vol. 6, no. 1, pp. 77–85, 2014.
    \item J. Zhang, P. Miao, and F. Zhang, “On optimal routing and spectrum allocation in elastic optical networks,” in 2023 2nd International Conference on Big Data, Information and Computer Network (BDICN). IEEE, 2023, pp. 284–287.
    \item S. Behera, J. George, and G. Das, “Effect of transmission impairments in co-ofdm based elastic optical network design,” Computer Networks, vol. 144, pp. 242–253, 2018.
    \item N. Amaya, G. Zervas, and D. Simeonidou, “Introducing node architecture flexibility for elastic optical networks,” Journal of Optical Communications and Networking, vol. 5, no. 6, pp. 593–608, 2013.
\end{enumerate}

}

\ifCLASSOPTIONcaptionsoff
  \newpage
\fi

\end{document}